\documentclass[prl,aps,twocolumn]{revtex4}
\usepackage{epsfig}
\usepackage{latexsym}

\newcommand{\note}[1]{{\tt #1}}
\renewcommand{\note}[1]{{}}

\begin{document}

\title{Manipulation and removal of defects in spontaneous optical patterns}

\author{R. Neubecker}
\affiliation{Institute of Applied Physics,
  Darmstadt University of Technology,
  Hochschulstr.~6, 64289~Darmstadt, Germany 
  }
\author{E. Benkler}
\affiliation{Institute of Applied Physics,
  Darmstadt University of Technology,
  Hochschulstr.~6, 64289~Darmstadt, Germany }
\author{R. Martin} 
\affiliation{Department of Physics, 
University of Strathclyde, 107 Rottenrow, Glasgow, G4 0NG, U.K.}
\author{G.-L. Oppo}
\affiliation{Department of Physics, 
University of Strathclyde, 107 Rottenrow, Glasgow, G4 0NG, U.K.}

\begin{abstract}
Defects play an important role in a number of fields dealing with
ordered structures. They are often described in terms of their topology, 
mutual interaction and their statistical characteristics. 
We demonstrate theoretically and experimentally the 
possibility of an active manipulation and removal of defects. 
We focus on the spontaneous formation of two-dimensional spatial 
structures in a nonlinear optical system, a liquid crystal light valve under
single optical feedback.
With increasing distance from threshold, the spontaneously formed hexagonal 
pattern becomes disordered and contains several defects. 
A scheme based on Fourier filtering allows us to remove defects and 
to restore spatial order. 
Starting without control, the controlled area is progressively expanded, 
such that defects are swept out of the active area. 
\end{abstract}

\maketitle

Defects are local deviations from a given ordered structure 
and have attracted a large interest in many areas of physics \cite{mermin}. 
The most prominent example is condensed matter physics \cite{chaikin,vanbueren}. 
This extends to adjacent areas, like spin lattices, liquid 
crystals \cite{leshouches}, but also to spontaneous structure formation, 
e.g.\ convection patterns \cite{convect}.

Defects have been catalogued in different classes depending on their topology
\cite{mermin}. Here we focus our attention on point-defects in
two dimensions, known as dislocations. 
These break discrete symmetry of the periodic structure locally. 
Dislocations are robust entities which 
cannot be removed by smooth local alterations of the pattern.
They can only be created or annihilated in pairs and correspondingly 
they are assigned a {\em topological charge} \cite{mermin}.

The spatial order that we consider here develops spontaneously through 
self-organisation and is studied in many branches of 
science \cite{crosshohmanne}. 
Spontaneous optical pattern formation is observed in numerous 
nonlinear media and optical architectures \cite{nlopatt}. 
For low input energy and low external stress, spatial structures evolve 
to ordered patterns while under increase of the stress parameter, spatial 
and/or spatio-temporal disorder 
({\em optical turbulence}) sets in. In certain instances such transitions have 
been shown to be mediated by the appearance and annihilation of defects 
\cite{crosshohmanne,defmed}.

In this communication, we deal with spatial disorder due to the presence 
of stable defects of the dislocation type. These configurations are 
often undesired since order maximises spatial correlations. We then describe 
a technique to progressively remove stable optical defects for a Kerr-Type 
nonlinearity in a so-called {\em single-feedback} setup \cite{singlefeedb}. 
Our approach should be realizable in other kinds of nonlinear optical devices 
and experiments, as well as in chemistry or in fluid convection, where 
symmetry-broken patterns are also observed.

In our feedback system the uniform state becomes modulationally
unstable at a given threshold of the laser (pump) intensity,
leading to the formation of stationary hexagonal patterns.
Above a second threshold of the pump intensity, 
patterns break up and become increasingly disordered, with defects 
appearing at random locations of the active area \cite{dyniap,dynelse}. 
The presence of spatial inhomogeneities can lower this second instability 
threshold.

The aim of this communication is to demonstrate the possibility to actively
(re)move and manipulate these defects. Since removing defects is equivalent to
the restoration of the ordered state, we need to apply a technique to control 
spatiotemporal disorder. 
In the regime of spontaneous disorder, the hexagonal state survives, but is unstable or 
marginally stable.
Our scheme of control allows us to select this particular solution out of 
the infinite set of coexisting (dynamical) solutions and stabilize it.  

\begin{figure}
\epsfxsize=\columnwidth \centerline{\epsffile{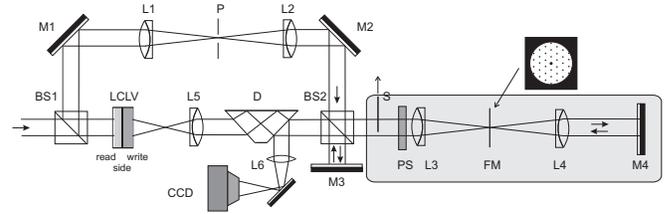}}
\caption{Figure 1: Scheme of the experimental setup, details are
    given in the text.
} \label{defsetup}
\end{figure}

Generalizing the control scheme introduced in \cite{strathcontr}, 
we discourage the growth of structures which do not belong to the desired, 
well-ordered {\em target state}. 
This is achieved by adding a feedback (control) loop, realized in an all-optical manner.
In this control loop, a spatial Fourier filter blocks the discrete number of modes
which constitute the spatially periodic {\em target pattern}. The remaining {\em
control signal} is fed back negatively into the system.  Consequently,
the system is driven towards the target state and the control signal
vanishes when the ordered state is reached.  This scheme has first been
proposed and been tested numerically \cite{strathcontr}, but has also
been proven to work in several experiments \cite{contrexp,SDcontr}. 

To manipulate defects in patterns, the control scheme described above
is modified. The intention is not to entirely replace
a disordered pattern by a regular one at once. Instead, the control is applied to 
a gradually increasing part of the spontaneous structure.
Since defects are topologically robust objects, they do not
vanish instantaneously. Instead, they can annihilate with corresponding partners
or move towards regions where the control
signal is not operating. 
Consequently, defects should be swept out by the front
separating the controlled, ordered part of the pattern to the uncontrolled
and disordered spatial region.

\begin{figure}
  \epsfxsize=80mm \centerline{ \epsffile{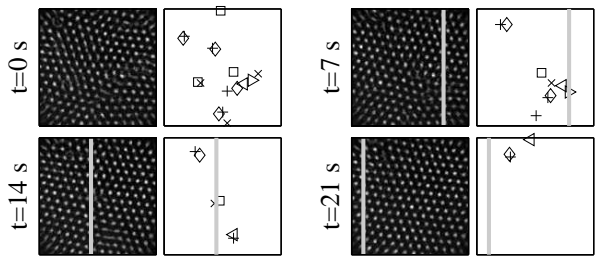}} \caption{
  Experimental sweeping of the pattern defects. 
  Left hand column: snapshots of the intensity distributions.
  The gray line indicates the position of the front separating the
  controlled from the uncontrolled part.  
  Right hand column: defects locations as extracted from the recorded
  images. The different marker types correspond to the hexagon modes. 
  $\Box$, $+$, and $\triangleleft$ indicate one charge,
  $\diamond$, {\sf x}, and $\triangleright$ indicate the opposite topological
  charge, respectively.}
  \label{pdemo3}
\end{figure}

In our single-feedback experiment, the optical Kerr-type nonlinearity 
is provided by a so-called Liquid Crystal Light Valve (LCLV). 
This device has an intensity-sensitive {\em write side}
(photoconductor layer) and a reflective {\em read side} with variable refractive
index (liquid crystal layer). Both are coupled by an external electric
ac field, applied by transparent electrodes. An intensity profile 
at the write side is transformed into a corresponding phase
profile of a light wave, which is reflected by the LCLV read side \cite{lclvpatt}.

The LCLV is put into a feedback loop: a uniform laser beam is phase
modulated and reflected by the LCLV read side. The modulated beam is then
fed back to the LCLV write side. It thereby propagates freely over a given
distance, which causes the spatial phase modulations to be transformed
into an intensity modulation.  In this way the feedback is closed and the beam
inside the loop modulates itself.
LCLV feedback systems in different configurations are successfully used for 
the investigation of spontaneous optical patterns \cite{lclvpatt,hyasia,dynelse,dyniap}.
In our particular realization, pattern formation is based solely on diffractional coupling.

The experimental setup is presented in Fig.~\ref{defsetup}.
A cw-Nd:YAG laser ($\lambda=532$~nm) acts as light source (not shown in the scheme).
The laser beam is phase modulated and reflected by the LCLV, and then guided to the
write side by means of beam splitters (BS), mirrors (M) and lenses. 
The lenses L1, L2 and the aperture P form a spatial
low pass filter. The dove prism D accounts for the correction of residual misalignments.
The transverse intensity distribution is recorded with a camera (CCD).

The shaded box in Fig.~\ref{defsetup} contains the control loop.
A fraction of the light wave is coupled out by beam splitter BS2. 
It passes the Fourier filter (4f-arrangement of the lenses L3, L4),
containing a mask FM in the Fourier plane, which blocks the target modes. The
remaining wave is reflected by mirror M4 and, after passing the filter
a second time, is reinjected into the system. The phase of this control wave is
adjusted by the phase shifter PS in order to achieve destructive interference,
i.e.\ negative feedback of the control signal. 
The space mask S allows the selection of the area to which this control is applied. 
It is located in a plane, which is imaged onto the LCLV write side.

\begin{figure}
  \centerline{ \epsfxsize=27mm  \epsffile{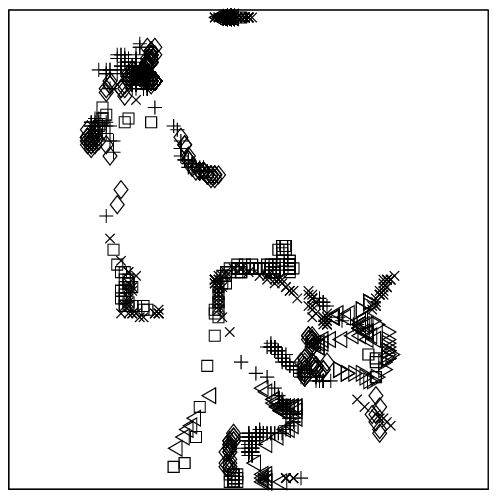} \hspace{5mm}
    \epsfxsize=27mm \epsffile{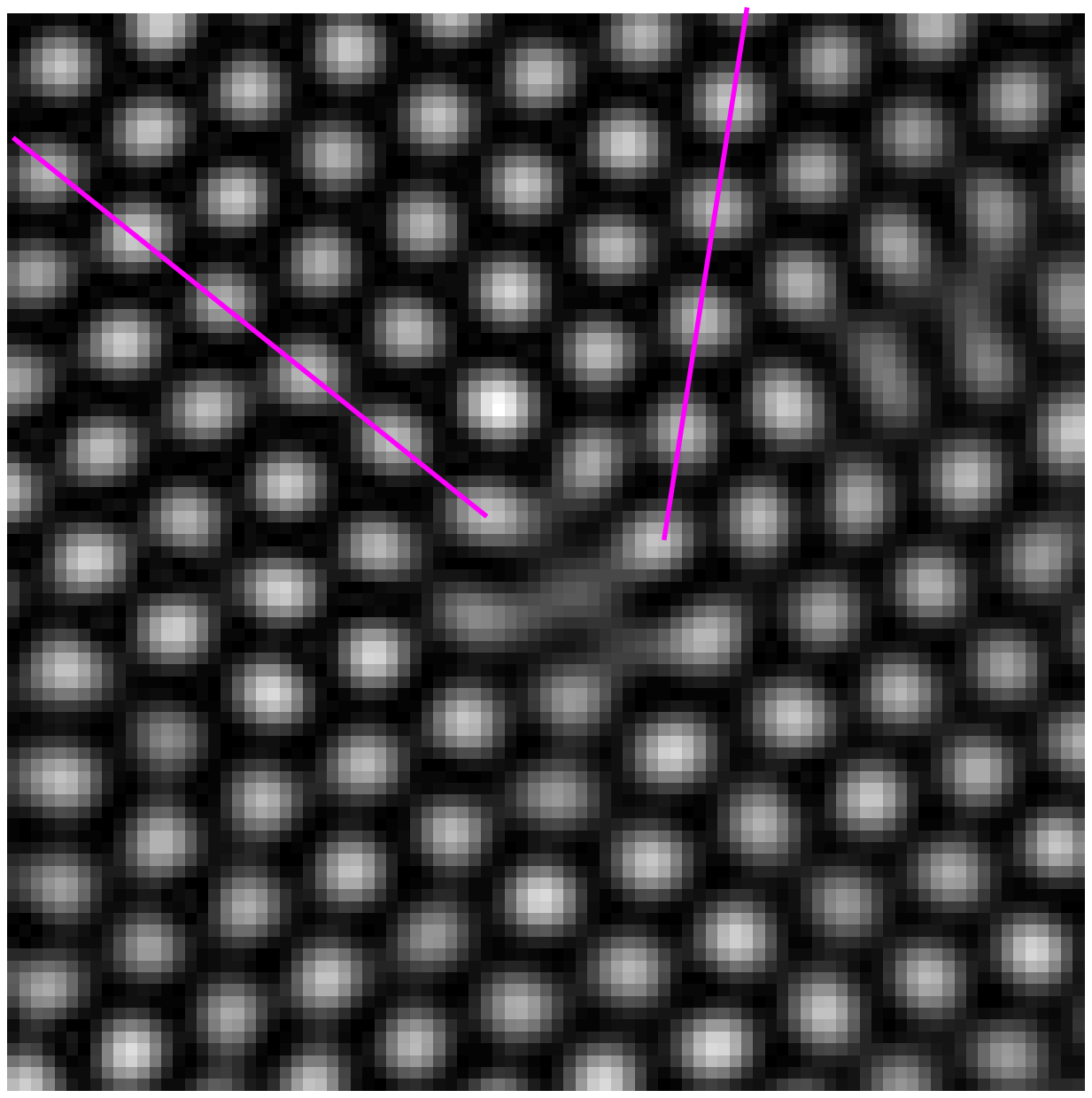} }
    \caption{Left panel: all defects locations, detected during the sequence of the moving
         control front displayed in Fig.~\ref{pdemo3}.
         Right panel: Experimental example of a penta-hepta defect, where two rolls 
         of different modes end close to each other (as indicated).}
  \label{alldef}
\end{figure}

For the experiment described here, the spatial low pass filter (L1,L2,P)
is used to block all wave numbers above the first critical band. 
This renders the transition to spatial disorder smoother
and makes it easier to choose the amount of disorder contained in
the spontaneous structure \cite{dyniap}. The pump intensity was set to
about six times the pattern threshold
intensity. In the uncontrolled system, a distorted hexagonal pattern is
observed (see Fig.~\ref{pdemo3}, first image at $t=0$), 
showing a slow dynamics on a time scale of approximately some hundred 
milliseconds. The structure consists of several ordered domains, between which 
single or strings of defects are located.
Care was taken to match the orientation of
the target pattern with the spontaneous pattern in
the region where the control is first applied.

After the system has reached an asymptotic disordered state, the space 
mask S is opened gradually from the right to the left side. Snapshots of the 
recorded sequence are shown in Fig.~\ref{pdemo3} where the front between 
controlled and uncontrolled regions is indicated by a gray line.  Eye inspection
of Fig.~\ref{pdemo3} already shows how hexagonal order is restored.

The ordering effect of our procedure becomes even more apparent when we extract
the defects from the recorded images (r.h.s.\ column of Fig.~\ref{pdemo3}).  
A hexagonal pattern consists of three independent Fourier modes, each
corresponding to a stripe pattern. Each of the stripe patterns can show
defects in the form of dislocations, i.e.\ of stripes ending somewhere
in the pattern.  Depending on the direction of the ending stripe,
the defect is assigned a positive or negative topological charge.
Consequently, a hexagonal pattern can contain six different types
of defects, corresponding to the six possible Burgers vectors of unit length
\cite{chaikin,vanbueren}. 
As long as we restrict the analysis to the topological properties only, 
domain boundaries can be regarded as strings of dislocations \cite{vanbueren}.

With the exception of few defects which remain 
anchored to local inhomogeneities, the defects are indeed swept
to the l.h.s.\ by the control signal, and are finally pushed out of the
active area. During this motion, mutual annihilation of pairs of
defects takes place as well as the temporary creation of few new defect pairs.

The effect of the control appears not to be strictly limited to the
r.h.s.\ part of the pattern that is directly exposed to the control
signal.  The number of defects in the uncontrolled area significantly
decreases even when the control front is relatively far away. This
can be associated to the long range spatial coupling:  the hexagonal
pattern is self sustained by diffractional coupling throughout its extent.
The restored order in the controlled part now promotes
order even in the uncontrolled, disordered part, causing a reduction of
the number of defects there.

\begin{figure}
\epsfxsize= \columnwidth \centerline{\epsffile{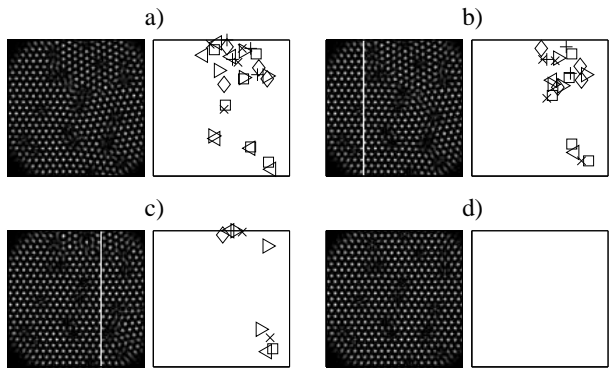}}
\caption{Numerical simulation of progressive removal of defects from 
an optical beam. The vertical bar signals the position of
the control mask moving from left to right. Top row panels corresponds 
to time $t=0$, (b) to $t=28$, (c) to $t=44$, and (d) to $t=80$ where 
$t$ has been normalised by the LCLV response time. Bottom row panels 
show the position of the defects}
\label{simu}
\end{figure}

We also observe that defects belonging to different modes of the hexagonal
pattern appear to stay close to each other.  This becomes evident by plotting 
all detected defect locations of the sequence into a 
single plot. Fig.~\ref{alldef} illustrates that the defects have
moved on definite paths.

In the investigation of model systems, Pismen, Tsimring et al.\  \cite{pistsim} have 
shown that two dislocations of different modes attract each other and tend 
to form bound states, so-called penta-hepta defects. 
An example of an experimentally observed penta-hepta defect is shown in Fig.~\ref{alldef}.
According to \cite{pistsim}, attracting as well as repelling forces can 
exist between different penta-hepta defects and the motion of such defects
depends very much on the embedding spatial structure.
Hence, we infer that the inhomogeneities in our experiment strongly influence the defect motion.

Numerical simulations of the progressive removal of defects in a
spatially disordered stable structure have been implemented by using
the LCLV model originally introduced in \cite{hyasia} and used later
in \cite{SDcontr}. 
All parameters of the fully quantitative model have been chosen in accordance
to the experiment.
Simulations above the pattern forming threshold, however,
did not reach stable spatially disordered configurations.  

In order to represent more faithfully the experimental observations, 
we have simulated local imperfections and inhomogeneities as
localised dips in the pump laser intensity.  These local imperfections
were of variable number, location, size and depth. A large variety
of stable or long term metastable disordered configurations formed by
patches of hexagonal patterns were obtained up to three times above
threshold. Remaining differences to the experiment, where metastable
disordered structures were observed higher above threshold
is easily explained by the simplified modelling of the inhomogeneities. 
With random initial conditions, a variety of spatially disordered structures 
were found, where defects tend to anchor in the vicinity of the local inhomogeneities.
Fig.~\ref{simu}(a) shows a stable (or long term metastable) disordered
structure observed three times above pattern formation threshold. 

The Fourier control scheme was implemented numerically in a way that the
area where control acts on the spontaneous structure can be chosen. 
Fig.~\ref{simu}(b) and (c) show the progressive sweeping
of the defects while the domain which corresponds to the target state
grows to finally occupy the entire area. 
In spite of the presence of pronounced local inhomogeneities, an ordered
hexagonal structure is restored and stabilised at the end of the
procedure (see Fig.~\ref{simu}(d)).

In agreement with the experimental results, we observe in the 
numerics progressive annihilation of dislocations ahead of the moving 
control front. In simulations with strong local inhomeogeneities
(like the ones reported here) we also observed a lag between the 
moving front and temporally surviving defects. This, again, is in 
agreement with experimental measures.

In conclusion, we have shown how defects can be swept out of spontaneous optical 
patterns by using a control technique based on Fourier filtering. 
The control is applied to a varying part 
of the spontaneously formed, slightly disordered hexagons.
Agreement is found between experiments on a LCLV single-feedback and numerical simulations 
on the full system model. We emphasized the important role played by local inhomogeneities in 
stabilizing spatial disorder by pinning the defects. 
This disorganising effect of inhomogeneities, which are difficult to eliminate in 
real experiments, can be overcome by our control technique.

Our control technique is independent of the nature of the nonlinearity 
and has been applied successfully to stabilise unstable patterns in 
models of saturable absorbers, lasers \cite{strathcontr}  and 
optical parametric oscillators (OPO) \cite{scott}.  
Presently, we are applying and modifying our elimination technique to 
systems with defects other than pattern dislocations. These include, 
for example, domain walls and vectorial defects in models of  
type-I and type-II optical parametric 
oscillators \cite{santaOSF}.

We expect our control scheme to work on a variety of experiments in 
optics and other physical systems such as extended chemical reactions 
and fluids, where stable defects have been observed.
Candidates for an all-optical implementation are photo-sensitive chemical systems,
which can be detected optically and have already have successfully been 
manipulated by imposed, spatially varying optical signals \cite{optchem}.
Otherwise, a digital realization with Fast Fourier transforms may well
be fast enough to allow to control systems with time constants in the range of milliseconds.

We acknowledge useful discussions with G.~K.~Harkness, 
and support by T.~Tschudi and by M.~Kreuzer. 
This research was partially funded by EPSRC (grants M19727, M31880, R04096), 
by SHEFC (grants VIDEOS and VISION), by the German science foundation (DFG 
grant SFB~185), and the Federal Ministry of Education and Research (BMBF 
grant 13N7311/8).
The collaboration between Strathclyde and Darmstadt was supported by the British 
Council and the DAAD in the framework of the ARC-programme.
G-L.O.\ acknowledges kind support from SGI.


\begin{thebibliography}{99}

\bibitem{mermin}
  N.D. Mermin, {\it Rev. Mod. Phys.} {\bf 51}, 591 (1979).
\bibitem{chaikin}
  P.~M.~Chaikin, T.~C.~Lubensky,
    {\em Principles of condensed matter physics},
    (Cambridge University Press, 1995)
\bibitem{vanbueren}
  H.~G.~van Bueren,
    {\it Imperfections in crystals}
    (North Holland Publishing, Amsterdam 1960)
\bibitem{leshouches}
  R.~Balian, M.~Kleman, and J.-P.~Poirier (eds.),
    Physics of Defects,
    Les Houches Lecture Series, Session XXXV, North-Holland Publishing
    (Amsterdam 1981)
\bibitem{convect}
  G.~Goren, I.~Procaccia, S.~Rasenat, V.~Steinberg,
    \prl {\bf 63}, 1237 (1989);
  P.~Cerisier, S.~Rahal, N.~Rivier,
    \pre {\bf 54}, 5086 (1996)
\bibitem{crosshohmanne}
  M.~C.~Cross and P.~C.~Hohenberg,
    {\it Rev. Mod. Phys.} {\bf 65}, 851 (1993);
  P.~Manneville,
    {\it Dissipative Structure And Weak Turbulence}
    (Academic Press, San Diego 1990)
%
\bibitem{nlopatt}
  N.~B.~Abraham and W.~J.~Firth eds.,
    \josab {\bf 7} (6,7) (1990);
  L.~A.~Lugiato ed.,
    Chaos, Solitons \& Fractals {\bf 4} (1994);
  L.~A.~Lugiato, M.~Brambilla, and A.~Gatti,
    Adv.\ Atom.\ Molec.\ Opt.\ Phys.\ {\bf 40}, 229 (1998);
  R.~Neubecker and T.~Tschudi eds.,
    Chaos, Solitons \& Fractals {\bf 10} (4,5), (1999);
  L.~A.~Lugiato, M.~Brambilla, A.~Gatti eds.,
    Adv.~At.~Mol.~Phys. {\bf 40}, 229 (1998);
  F.~T.~Arecchi, S.~Bocaletti, P.~L.~Ramzza,
    Phys.~Rep.\ {\bf 318}, 83 (1999).
%
\bibitem{defmed}  
  P.~Coullet, L.~Gil, J.~Lega,
    Phys.~Rev.~Lett.\ {\bf 62}, 1619 (1989)
%
\bibitem{singlefeedb}
      G.~D'Alessandro and W.~J.~Firth,
        \pra {\bf 46}, 537 (1992);
      R.~Macdonald and H.~J.~Eichler,
        Opt.Commun.\ {\bf 89}, 289 (1992);
      M.~Tamburrini, M.~Bonavita, S.~Wabnitz, and E.~Santamato,
        Opt.Lett. {\bf 18}, 855 (1993);
      G.~Grynberg, A.~Ma\^{i}tre, A.~Petrossian,
        \prl {\bf 72}, 2379 (1994);
      T.~Ackemann, Yu.~A.~Logvin, A.~Heuer, and W.~Lange,
        \prl {\bf 75}, 3450 (1995);
      T.~Honda, H.~Matsumoto, M.~Sedlatschek, C.~Denz, and T.~Tschudi,
        Opt.Commun. {\bf 33}, 293 (1997)
%
\bibitem{dyniap}
  R.~Neubecker, B.~Th\"ur\-ing, M.~Kreu\-zer, and T.~Tschu\-di,
    Chaos, Solitons \& Fractals {\bf 10}, 681 (1999);
  G.Schliecker, R.Neubecker,
    Phys.~Rev. E {\bf 61}, R997 (2000).
%
\bibitem{dynelse} 
  M.~A.~Vorontsov, J.~C.~Ricklin, G.~W.~Carhart,
    Opt.~Eng.\ {\bf 34}, 3229 (1995);
  P.~L.~Ramazza, S.~Ducci, S.~Boccaletti, and F.~T.~Arecchi,
    J.~Opt.~B: Quant.~Semiclass.~Opt.\ {\bf 2}, 399 (2000).
%
\bibitem{strathcontr}
  R.~Martin, A.~J.~Scroggie, G.-L.~Oppo, and W.~J.~Firth,
    \prl {\bf 77}, 4007 (1996);
  G.~K.~Harkness, R.~Martin, A.~J.~Scroggie, G.-L.~Oppo,
    and W.~J.~Firth,
    \pra {\bf 58}, 2577 (1998);
%
\bibitem{contrexp}
  A.~V.~Mamaev and M.~Saffman,
    \prl {\bf 80}, 3499 (1998);
  S.~J.~Jensen, M.~Schwab, and C.~Denz,
    \prl {\bf 81}, 1614 (1998);
  T.~Ackemann, B.~Giese, B.~Sch\"apers, and W.~Lange,
    J.\ Opt.\ B {\bf 1}, 70 (1999);
  E.~Benkler, M.~Kreuzer, R.~Neubecker, and T.~Tschudi,
    \prl {\bf 84}, 879 (2000);
  R.~Neubecker, E.~Benkler,
    Phys.~Rev.~E {\bf 65}, 066206 (2002).
%
\bibitem{SDcontr}
  G.~K.~Harkness, G.-L.~Oppo, E.~Ben\-kler, M.~Kreuzer,
    R.~Neubecker, and T.~Tschudi,
    J.\ Opt.\ B {\bf 1}, 177 (1999)
%
\bibitem{lclvpatt}
      B.~Th\"uring, R.~Neubecker, and T.~Tschudi,
        Opt.\ Commun.\ {\bf 102}, 111 (1993);
      E.~Pampaloni, S.~Residori, and F.T.~Arecchi,
        Europhys.~Lett. {\bf 24}, 647 (1993);
      M.~A.~Vorontsov and W.~J.~Firth,
        \pra {\bf 49}, 2891 (1994);
      F.~T.~Arecchi, S.~Boccaletti, S.~Ducci, E.~Pampaloni, P.-L.~Ramazza,
        and S.~Residori,
        J.~of Nonlin.~Opt.~Phys.~and~Mat. {\bf 9}, 183 (2000)
%
\bibitem{hyasia}
  R.~Neubecker, G.-L.~Oppo, B.~Th\"{u}ring, and T.~Tschudi,
    \pra {\bf 52}, 791 (1995);
  B.~Th\"uring, R.~Neubecker, M.~Kreu\-zer, E.~Ben\-kler,
    and T.~Tschu\-di,
    Asian J.~Phys.\ {\bf 7}, 453 (1998)
%
\bibitem{pistsim}
  L.~M.~Pismen, A.~A.~Nepomnyashchy,
    Europhys.~Lett.\ {\bf 24}, 461 (1993);
  M.~I.~Rabinovich, L.~S.~Tsimring,
    Phys.~Rev.~E {\bf 49}, R35 (1994);
  L.~S.~Tsimring,
    Phys.~Rev.~Lett.\ {\bf 74}, 4201 (1995).
%
\bibitem{scott}
  S. Sincalir, PhD thesis, University of Strathclyde (2002)
%
\bibitem{santaOSF}
G.-L. Oppo, A.J. Scroggie and W.J. Firth, Phys. Rev. E {\bf 63}, 066209 (2001);
M. Santagiustina, E. Hernadez-Garcia, M. San Miguel, A.J. Scorggie, and 
G.-L. Oppo, Phys. Rev. E {\bf 65}, 036610 (2002).
%
\bibitem{optchem}
    F.~Fecher {\em et al.},
     Chem.~Phys.~Lett {\bf 313}, 205 (1999);
    M.~Dolnik, I.~Berenstein, A.~M.~Zhabotinsky, and I.~R.~Epstein,
      \prl {\bf 87}, 238301 (2001).
      

\end{thebibliography}
\end{document}